\documentclass{osa-article}

\journal{osajournal}

\articletype{Research Article}

\usepackage{amsmath}
\begin{document}

\title{Design principles for >90\% efficiency and >99\% indistinguishability broadband quantum dot cavities}

\author{David Dlaka,\authormark{1,*} Petros Androvitsaneas,\authormark{1,2} Andrew Young,\authormark{1}   Qirui Ma,\authormark{1} Edmund Harbord,\authormark{1} and Ruth Oulton\authormark{1}}

\address{\authormark{1}Quantum Engineering Technology Labs, H. H. Wills Physics Laboratory and
Department of Electrical and Electronic Engineering, University of Bristol, BS8 1FD, UK\\
\authormark{2}School of Engineering, Queen's Buildings,
Cardiff University, Cardiff CF24 3AA, United Kingdom}

\email{\authormark{*}david.dlaka@bristol.ac.uk} 



\begin{abstract}

Quantum dots have the potential to be the brightest deterministic single photon source with plausible high end applications in quantum computing and cluster state generation. In this work, we re-examine the design of simple micropillars by meticulously examining the structural effects of the decay into leaky channels beyond the atom-like cavity estimation. We show that precise control of the side losses with the diameter and avoidance of propagating Bloch modes in the DBR structure can result in easy to manufacture broadband (Q$\approx750-2500$) micropillars and demonstrate extremely high internal efficiency ($90.5\%-96.4\%$). We also demonstrate that such cavities naturally decouple from the phonon sideband, with the phonon sideband reducing by a factor of $5-33$ allowing us to predict that the photons should show $99.2\%-99.8\%$ indistinguishability. \\ 
\end{abstract}

\section{Introduction}
Highly efficient sources of quantum light - single and entangled photons - that are readily manufactured, are vital for quantum technologies and quantum communications. Quantum dots (QDs) have near unity internal quantum efficiency and their utility extends well past single- or pair- photon sources as they can be used as light-matter interfaces with a possible perfect non-linearity, which requires a simultaneous near-perfect indistinguishability and efficiency. To meet these requirements, one exploits cavity quantum electrodynamics (cQED) to create photonic structures, such as a micropillar cavity, which efficiently funnels the light into a well-defined optical mode.
The micropillar single photon source (SPS) consists of a quantum dot placed inside a resonator made up of a $\lambda/n$ "cavity" layer and a quarter-wavelength distributed Bragg reflector (DBR) stack on either end of the cavity. The resulting emitter-cavity interaction results in enhancement of emission along the growth axis, and by etching the sides of the planar structure into cylindrical micropillars, the field at the dielectric interfaces is laterally confined therefore reducing the resonator leakage through the sides and increasing decay into favourable channels.

Research into solid state single photon devices has produced very promising micropillar-quantum dot emitters. The true potential of any cavity-enhanced SPS only occurs if the QD and cavity are degenerate, leading to various techniques being used to bring the QD energy into resonance with the cavity. Notably efficient devices had been experimentally realised, applying external E-field via a bias potential across electrodes on the cavity to Stark-shift the QD into resonance with the cavity, showing 65\% efficient resonance fluorescence in high-Q (12000) pillars \cite{Somaschi2016}. Similar work in spectrally narrow cavities showed $\approx 66\%$ collection into 0.65NA with simultaneously high single-photon purity and indistinguishiability \cite{Ding2016}. Further experimental demonstrations of bright micropillars was reported with $74.4 \pm 4\%$ overall efficiency at IR wavelengths\cite{Unsleber2016}. More recently, novel approaches such as planar (non-etched) half-cavities with a microlens mirror completing the resonator achieved 57\% end-to-end efficiency through a single mode fiber \cite{Tomm2021}.

A common drawback most current state-of-the art micropillar devices share is the reliance on strong Purcell enhancement which is achieved by resorting to narrow-band resonators, with Q-factors $>10000$ and resulting cavity linewidths $< 0.1 meV$, which is too narrow to fit the short optical pulses (1-2 ps) required to avoid multiple excitation-relaxation cycles which are catastrophic to the single photon purity~\cite{Fischer2017}. High-Q pillars are extremely sensitive to breaks in the degeneracy of the emitter-resonator system, leading to a particular vulnerability to fabrication tolerances where asymmetry in the cylinder (ellipticity) causes the resonator to split into two non-degenerate H and V $\Gamma_\text{Cavity}$ modes. With this in mind, broadband cavities have attracted attention and research efforts have demonstrated that low-Q cavities can facilitate deterministic light-matter interactions, the efficiency of which can be maximised by suppressing lossy modes\cite{Androvitsaneas2016, PetrosACS19},  \cite{Gines2022}. 

Other designs include exotic micropillar variations such as nanotrumpets that have been lifted off and placed on gold mirrors predicting 75\% first-lens efficiency \cite{Munsch2013}, or hourglass-like micropillars which predicts strong coupling between a spatially and spectrally ideal emitter, though in very narrow cavity (Q > 30k) \cite{Osterkryger2019,Gaal2022}. Other techniques to utilise micropillars as on-demand single photon sources include the introduction of controlled ellipticity in the cylinder structure, where the break in the radial symmetry splits the cavity decay channel into non-degenerate modes allowing for the pumping into one and extraction from the other, thus removing the cross-polarisation collection limit of more rudimentary pumping techniques, with efficiencies as high as 60\% in structures at $Q=4000$\cite{Wang2019} and potentially 89\% for $Q=25000$\cite{Gur2021} have been reported. However, asymmetric cavity designs preclude the encoding polarisation information in the qubit.

Furthermore, as these proposals involve technically difficult and resource intensive fabrication, there is still the necessity of a more scalable answer to the solid state SPS problem. In this work, we return to the basic micropillar structure due to its simpler manufacturability and proven potential as an ideal deterministic and versatile single photon source. By discussing phenomena which are applicable to any cavity, we present a universally applicable and simple design strategy to maximise the device efficiency, indistinguishability, and bandwidth by providing an in-depth analysis of the hitherto under-explored low-Q micropillars. We also consider the suppression of the phonon-photon interactions, leading to very high indistinguishability and a near elimination of the phonon sideband.

\section{Methodology}

\begin{figure}[ht!]
    	\center
    	\includegraphics*[width=0.75\linewidth]{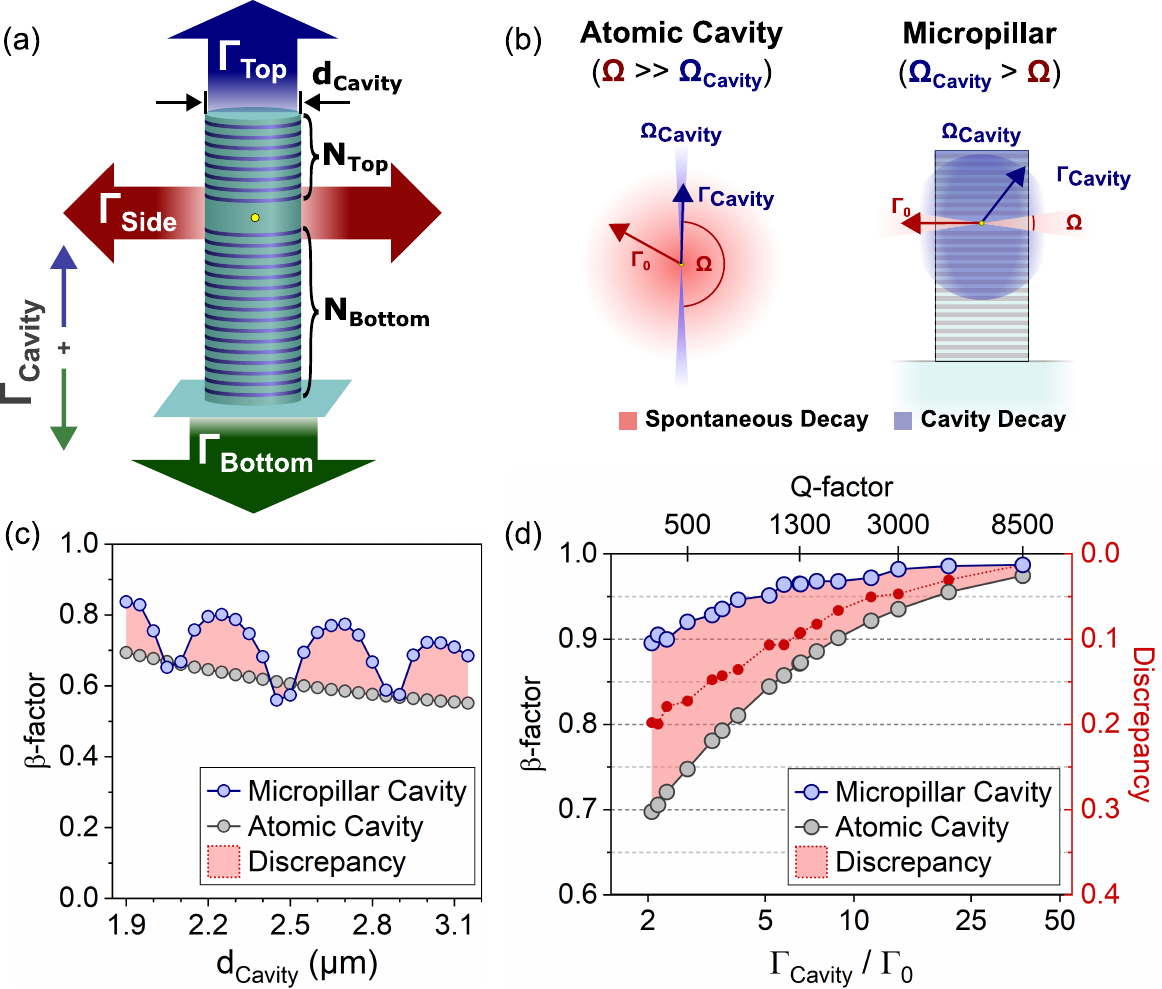}
    	\caption{(a) The micropillar diameter $d_\text{Cavity}$ and numbers of DBR layers in the stacks on top and bottom of the cavity ($N_\text{Top}$ and $N_\text{Bottom}$, respectively) are varied while the power decay through the top, side, and bottom faces of the micropillar is calculated using FDTD. (b) An atomic emitter with an infinitesimal solid angle subtended by the cavity $\Omega_\text{Cavity}$ is much less favourable than a micropillar where the cavity spatially dominates over the spontaneous emission into $\Omega$. (c) The calculated micropillar $\beta$-factor (blue) and the atom-like cavity approximation of the same, $\beta_\text{Atom}\approx F_P/(F_P+1)$ (grey) show distinct trends when $d_\text{Cavity}$ is varied against a BCB background. (d) The discrepancy (red, right axis) between the pillar (blue) and atomic (grey) resonators remains significant across most of the weak-coupling regime with >0.05 errors at $F_P < 25$, $Q<5000$.}
    	\label{AtomCavityFigure}
	\end{figure}

We use a commercially available FDTD solver \cite{Lumerical} to fully examine the micropillar cQED by probing the diameter and DBR stack effects on the efficiency using a 7/35.5, $d_\text{Cavity}=2.25\mu$m "pilot" pillar, and at the end we use these effects to design high-brightness broadband pillars. We use benzocyclobutene (BCB) as a background medium for some simulations, as indicated in the figure captions.We examine all decay rates through the pillar faces while making no assumptions about the side leakage or the background decay rate (Fig. \ref{AtomCavityFigure}(a)), but relying on the fundamental relation

\begin{equation}
        \Gamma_\text{Cavity}+\Gamma_\text{Side} = F_P \Gamma_0 \label{tPurcellEq}
\end{equation}

where $\Gamma_\text{Cavity}$ is the decay rate into the cavity, $\Gamma_\text{side}$ is the decay rate into leaky side modes and $F_P$ is the Purcell factor, describing the enhancement of the decay rate if the emitter was placed in bulk, $\Gamma_0$. The motivation is to avoid avoid making the common atom-like approximation that the cavity subtends a tiny angle, leading to the commonly used expression to parameterise the active cavity efficiency, $\beta_\text{Atom} = F_P/(F_P+1)$. A non-infinitesimally narrow cavity can have off-resonant Purcell \emph{suppression}, which is a crucial tool in maximising the device efficiency (Fig. \ref{AtomCavityFigure}(b)). Furthermore, the atomic cavity estimation above only considers the effects of the vertical resonator and ignores any effects arising from the finite diameter and reflections at the radial edges of the cavity. The failure to take this into account leads to errors where the real $\beta$-factor (the efficiency to couple the emitter to a particular cavity mode) and the approximated  one are very different (Fig. \ref{AtomCavityFigure}(c-d)). To avoid this, we look at the decay rates through the 3 pillar faces and work out our efficiency factors more thoroughly such that 

\begin{equation}
        \beta = \frac{\Gamma_\text{Cavity}}{\Gamma_\text{Total}} = \frac{\Gamma_\text{Cavity}}{\Gamma_\text{Cavity}+\Gamma_\text{Side}}\label{BetaEq}
    \end{equation} 

The $\beta$-factor describes the source-cavity coupling, and by using $\Gamma_\text{Cavity}$ and $\Gamma_\text{Side}$ rather than the Purcell factor, we take into consideration that a cavity which dominates the angular space demands less Purcell enhancement over the spontaneous decay to achieve high $\beta$-factors \cite{Carmichael2009}. For a full appreciation of the internal efficiency $\xi$, we must also consider the passive efficiency $\eta$ which quantifies the leakage of the cavity mode into the substrate such that

\begin{align}
    \eta &= \frac{\Gamma_\text{Top}}{\Gamma_\text{Cavity}} = \frac{\Gamma_\text{Top}}{\Gamma_\text{Top}+\Gamma_\text{Bottom}} \label{EtaEq} \\
    \xi &= \beta \eta = \frac{\Gamma_\text{Top}}{\Gamma_\text{Total}} \label{XiEq}
\end{align}

By varying the fundamental structural parameters (Fig. \ref{AtomCavityFigure}a) of the pilot device, we identify several competing decay channels in the micropillar and use this as a design guideline to produce an array of device designs. In this work we focus on 3D simulations of devices in the telecoms O-band, though analogous results observed for 910nm micropillars indicate that our optimisation method applies to any wavelength.  In quantifying the active coupling through the power decay rather than the Purcell factor, we avoid making the small angle cavity (atom-like) approximation; that is, in atomic emitters the solid angle subtended by the cavity can be approximated to be very small such that $\Omega \gg \Omega_\text{Cavity}$. This allows for the approximation that the k-space subtended by $\Omega_\text{Cavity}$ can be neglected and it can be estimated that $\Omega \approx 4\pi^2$, leading to the commonly used $\beta_\text{Atom}$ estimation. This ignores both the possibility of suppressing the decay rate into the non-cavity $\Omega$ as well as the waveguiding provided by the micropillar (Fig.\ref{AtomCavityFigure}b). 

By introducing an edge along the radial dimension, the source no longer "sees" an infinite cavity semiconductor in $\Omega$, but a reflective surfaces instead; the decay rate of the corresponding k-vectors is therefore \emph{not} $\Gamma_0$ anymore, but some other decay $\Gamma_\text{Side}$. This possibility is not captured by the small $\Omega_\text{Cavity}$ approximation as can be seen in Fig.\ref{AtomCavityFigure}c where the decay-based calculation of the pilot pillar $\beta$-factor behaves very differently to the Purcell factor estimation of the same. Even at a fixed diameter, approximating $\beta$ with the Purcell factor underestimates the true source-cavity active coupling throughout a significant range ($F_P\leq50$, $Q\leq8500$) of the weak coupling regime. This distinction between $\Gamma_0$ and $\Gamma_\text{Side}$ is important to highlight, as the large solid angle subtended by the cavity allows for the possibility of an off-resonant density of states which is lower than the Einstein A coefficient, leading to cavity-inhibited spontaneous emission into the side modes. This non-trivial detail is crucial in optimising both low- and high-Q micropillars as $\Gamma_\text{Side}$ will need to be minimised for near-perfect efficiency.

\section{Side-loss Suppression}

In Fig.\ref{AtomCavityFigure}c, there is a clear oscillation of the $\beta$-factor with diameter; we have observed this behaviour in previous FDTD simulations having found it to be in agreement with experimental measurements \cite{Gines2022}. It has also been observed with analytical examinations using the Fourier Modal Method \cite{Wang2021}, where a detailed treatment describes this periodicity as a result of guided modes along a fibre-like model of the pillar. We build on this work by implementing the numerical FDTD technique which doesn't require the simplification of the vertical DBR pillar structure, verifying the main result but showing that there are discrepancies when the index of the surrounding medium is changed. We therefore present an alternative 1-D Fabry Perot model which allows for a simple yet effective design strategy depending solely on the desired pillar wavelength and the cavity semiconductor material.

\begin{figure}[ht!]
    	\center
    	\includegraphics*[width=0.6\linewidth]{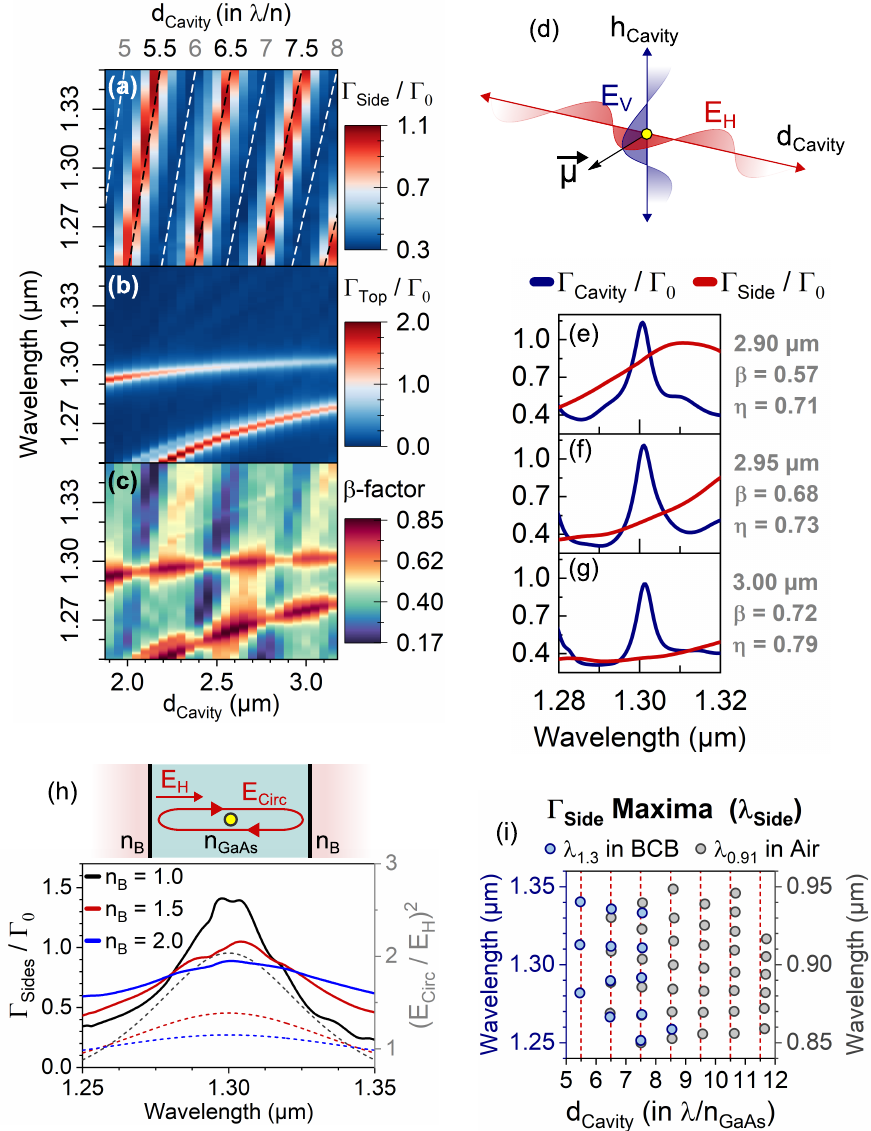}
    	\caption{(a) The decay rate through the sides of a micropillar in a BCB background medium, $\Gamma_\text{Sides}$ (in units of bulk decay rate $\Gamma_0$) is shown as a function of wavelength and micropillar diameter. The white dashed lines represent $d_\text{Cavity}$ values corresponding to full-wavelengths, whereas black dashed lines mark half-wavelength diameters. (b) The decay rate through the top of the micropillar. (c) The resulting $\beta$-factor. (d) The emitter dipole moment $\vec{\mu}$ is simultaneously perpendicular to vertical \emph{and} horizontal E-fields shown in blue and red, respectively. (e-g) The decay rates through the top (blue) and red (sides) normalised to the bulk decay rate $\Gamma_0$ for pillars with increasing diameters.   The decay rate through the top (blue) remains largely consistent while the side decay (red) is redshifted when $d_\text{Cavity}$ increases. (h) A plot of $\Gamma_\text{Side}/\Gamma_0$ for different surrounding dielectrics. The horizontal cavity Purcell enhancement and suppression of the side leakage grows weaker when the refractive index contrast between the cavity dielectric and the surrounding medium is reduced, and is accurately modelled as a 1-D Fabry-Perot resonator (dashed lines, right axis). (i) The resonant frequencies of leaky-side modes for O-band pillars immersed in BCB (blue, left axis) and 0.91$\mu$m pillars in air (grey, right axis) are independent of the surrounding medium and occur at half-wavelength cavity diameters (red dashed lines). The side maxima clearly overlap the half-wavelength dashed lines regardless of the background index, in contrast to the fibre-guided mode model.}
    	\label{DiameterModesFig}
	\end{figure}

In Fig.\ref{DiameterModesFig}a we calculate the power decay through the side, $\Gamma_\text{Side}$ normalised to the decay rate in bulk GaAs, $\Gamma_0$, as a function of wavelength and $d_\text{Cavity}$. It can be seen that the strong horizontal leakage occurs at diameters corresponding to half-wavelength values (black dashed lines) whereas at full-wavelength diameters (white dashed lines), $\Gamma_\text{Side} \ll \Gamma_0$. The lossy $\Gamma_\text{Side}$ can be suppressed from 1.1 to 0.3 $\Gamma_0$; as the fundamental cavity mode varies very little with diameter (Fig. \ref{DiameterModesFig}b), the resulting active coupling to the cavity as described by Eq. \eqref{BetaEq} and presented in Fig. \ref{DiameterModesFig}c shows regions of high $\beta$-factor ($\approx 85\%$) separated by areas of much lower $\beta$ ($\approx 55\%$) when the peaks in $\Gamma_\text{Side}$ spectrally overlap the $\Gamma_\text{Cavity}$ modes.

The diameter's effect on the side leakage is considerable as the etching introduces a reflective radial edge, resulting in a horizontal resonator formed by fields reflecting at the cavity-air boundary. The horizontal standing modes which are analogous to the vertical ones, result in distinct and separate horizontal modes in addition to the cavity modes. (Fig. \ref{DiameterModesFig} (d)). When the diameter is changed, the resonant frequencies of the horizontal resonator are shifted much like the cavity height controls the resonances of the vertical modes. 
In Fig. \ref{DiameterModesFig}(e-g), the cavity mode (blue) remains unperturbed by the small increments in the pillar diameter, while the much broader side mode (red) is shifted as the wider diameters means a longer horizontal cavity, and standing waves of a higher wavelength are resonant instead. While the $\Gamma_\text{Side}$ experiences Purcell enhancement on resonance, the non-atomic treatment of the micropillar allows us to see that the side leakage is Purcell suppressed off resonance, causing $\beta$ to increase from 57\% to 72\% as $d_\text{Cavity}$ goes from 2.9 to 3$\mu$m (the $\eta$-factor also changes - this is discussed in Section \ref{BlochSection}). By tuning the horizontal resonant frequencies away from the vertical cavity ones, we can use off-resonant Purcell suppression of the side channel to increase $\beta$ well past $\beta_\text{Atom}$ because $\Gamma_\text{Side} \ll \Gamma_0$. This suppression is the crucial effect that is not captured by the atomic estimation and is responsible for the results in Fig. 1(c-d).

To further investigate the suitability of the 1-D Fabry Perot model of the horizontal modes, we keep a constant pillar diameter and instead change the index of the background medium, $n_B$. In Fig.\ref{DiameterModesFig}(h) we show a diagram of the input field $E_H$ and the field reflected at the edge and circulating inside the resonator, $E_\text{Circ}$, as described by standard model in \cite{Ismail2016} where the resonator enhancement is given by

\begin{equation}
    \frac{|E_\text{Circ}|^2}{|E_\text{H}|^2} = \frac{1}{(1-R)^2}
    \label{ECircEq}
\end{equation}

 where R is the Fresnel coefficient given by $(n_\text{GaAs}-(n_\text{B}))^2/((n_\text{GaAs}+(n_\text{B}))^2$. The plot in Fig.\ref{DiameterModesFig}(h) shows how $\Gamma_\text{Side}$ is affected by the increase in $n_\text{B}$, whereas the dashed lines correspond to the right axis and show the resonator enhancement as predicted by Eq.\eqref{ECircEq}. The decreasing contrast weakens the reflection at the radial boundary, reducing the wavelength variation of the Purcell effect (both on-resonant enhancement, and off-resonant suppression) of the side resonator, but does not otherwise affect their spectral position. This is in contrast to the fibre-guided mode model, but fortunately this allows us to optimise the diameter using solely the desired micropillar wavelength and the material of the cavity layer i.e. for optimal $\beta$-factors, $d_\text{Cavity} = m\lambda/n_\text{Cavity}$ where $m \in N$.
 To confirm this, we examine the resonant wavelengths of the horizontal cavity for a range of diameters, wavelengths (1.3 and 0.91$\mu$m), and background materials (air and BCB). In Fig. \ref{DiameterModesFig}(i) this has been plotted and shows the independence of the side modes from the background index, and that for the 5-10 $\lambda/n$ range (corresponding to diameter 1.5-3.0$\mu$m) it is sufficiently accurate for use as a predictive tool. For larger diameters, a full 2-D treatment of the resonator using Bessel functions would be necessary for accuracy.
 
\section{Propagating Bloch Modes}
\label{BlochSection}
\begin{figure}[h!]
    	\center
    	\includegraphics[width=0.75\linewidth]{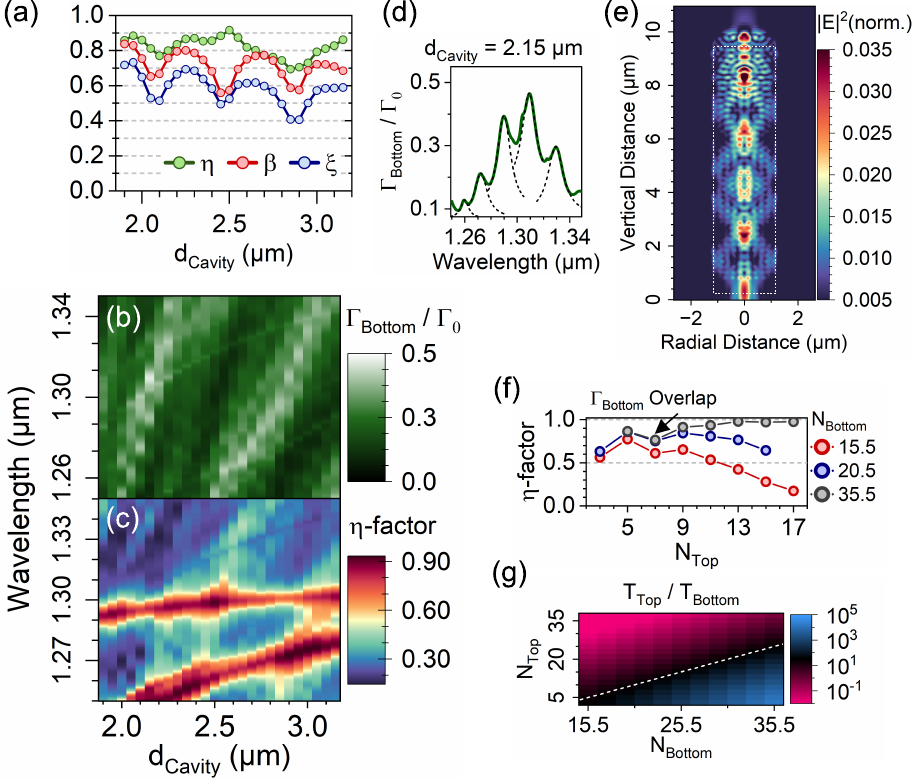}
    	\caption{(a) A plot of the on-resonance peak $\beta$ and $\eta$ factors of the 7/35 pilot pillar in BCB shows a non-identical periodicity in $d_\text{Cavity}$, affecting $\xi = \beta\eta$ (blue). (b) The decay rate through the bottom of the BCB encased pillar as a function of wavelength and diameter reveals a plethora of $d_\text{Cavity}$ dependent modes which propagate along the DBR structure and decay downward. (c) The $\eta$-factor is worsened where the bottom-leaky modes are degenerate with the cavity mode. (d) The bottom leakage $\Gamma_\text{Bottom}$ at $d_\text{Cavity}=2.15\mu$m shows many propagating Bloch modes, rendering this a poor diameter choice despite the high $\beta$-factor. (e) The E-field intensity along a plane bisecting the structure shows the Bloch modes propagating inside the structure, with intense spots forming where lower and higher (f) The $\eta$-factor of micropillars surrounded by air at the beta-optimal $d_\text{Cavity} = 2.25\mu$m ($6\lambda/n$) with various bottom DBR pairs $N_\text{Bottom}$ is shown as a function of $N_\text{Top}$; the marked performance dip is due to re-direction of the field into the substrate by propagating Bloch modes. (g) The Transfer Matrix Method (TMM)-calculated ratio of the transmissivity between the top ($T_\text{Top}$) and bottom ($T_\text{Bottom}$) mirrors shows values for $N_\text{Top},N_\text{Bottom}$ which produce strongly downward (pink) and upward (blue) pointing cavities.}
    	\label{BottomDecayFig}
	\end{figure}

While the active coupling $\beta$ quantifies the probability of Purcell enhanced decay into the cavity, it doesn't differentiate between emission vertically upwards or downwards. In order to get the full picture, we also examine the passive efficiency $\eta$ (Eq.\eqref{EtaEq}); it can be seen in Fig.\ref{DiameterModesFig}(e-g) that despite the 2.25$\mu$m and 2.7$\mu$m pillar having similar $\beta$-factors (0.80 and 0.77, respectively), differences in the passive efficiency $\eta$ (0.87 and 0.77) lead to strongly contrasting efficiencies $\xi$ (0.69 to 0.60). These efficiency losses arise from Bloch modes which can propagate through the structure and into the substrate despite the high reflectivity stop-band, providing a secondary decay route. In this section, we explore the dependence of the Bloch mode resonant wavelengths on the diameter $d_\text{Cavity}$ and the top and bottom DBR stack lengths (controlled by $N_\text{Top}, N_\text{Bottom}$) in Fig.\ref{BottomDecayFig}, demonstrating the crucial need to spectrally detune these lossy modes from the cavity mode and use off-resonant Purcell suppression to minimise any $\Gamma_\text{Bottom}$ leakage.

The need to expand our design strategy past the side loss minimising vertical-and-horizontal resonant cavity model is evident in the plot of the peak/on resonance efficiency factors as a function of $d_\text{Cavity}$ in Fig.\ref{BottomDecayFig}(a), where we observe variations in the $\eta$ factor which are independent of the $\beta$-factor ones and cause some full wavelength diameters to be much more efficient, i.e. have a higher $\xi$ than others. In Fig.\ref{BottomDecayFig}(b), we show the wavelength-resolved decay into the substrate, $\Gamma_\text{Bottom}$, as a function of the diameter where the absence of $\Gamma_\text{Cavity}$ contributions signals that the well guided modes of the cavity are reflected upward by the 35.5 DBR pairs in the bottom stack, while one can identify a variety of low- and high-Q modes whose resonant wavelength is affected by $d_\text{Cavity}$. As a result of the enhanced $\Gamma_\text{Bottom}$, it can be seen from the heatmap in Fig.\ref{BottomDecayFig}(c) of $\eta$ as a function of wavelength and diameter that the passive efficiency is high when the cavity mode dominates and diminished when $\Gamma_\text{Bottom}$ is spectrally degenerate. Unlike the $\beta$-factor and the single $\Gamma_\text{Side}$ mode and its harmonics, the peaks in $\Gamma_\text{Bottom}$ are comprised of a plurality of repeated modes as exampled in Fig.\ref{BottomDecayFig}(d), and a detuning strategy is not as straightforward.

The mechanism behind this energy transfer into the substrate had been observed in analysis of narrow micropillar structures \cite{Lalanne2004}, with analogous effects attributed to Fabry-Perot modes in photonics crystals preventing near-ideal photon extraction \cite{Rigal2018}.  A detailed analytical treatment in agreement with experimental measurements \cite{Lecamp2005} followed shortly, where it was identified that modes exist whose effective index is solely real, resulting in no Bragg reflection and unrestricted propagation through the micropillar structure. It was further shown in \cite{Lecamp2005} that the well-guided cavity mode $\Gamma_\text{Cavity}$ can scatter at a reflective interface and excite propagating Bloch mode, with the energy leakage rate proportional to the scattering strength of the interface (i.e. reflected field intensity) as well as the spatial mode overlap at said interface. An example of a micropillar with strong leaky Bloch modes is presented in Fig.\ref{BottomDecayFig}(e), showing the Fourier transform of $E(t)$ in a plane going through the pillar middle. The colormap is scaled to highlight the propagating Bloch modes which propagate through the structure; this particular micropillar is 7/35.5 at 2.25$\mu$m diameter and the highly reflective GaAs-air interface at the micropillar top occurs at a point of significant spatial overlap between lower- and higher-order Bloch modes. This leads to significant excitation of the Bloch modes, leading to $\eta$-factor reduction. 
Plots of the on-resonance passive efficiency of the 2.25$\mu$m pillars with 15.5, 20.5, and 35.5 bottom DBR pairs have been shown in Fig.\ref{BottomDecayFig}(f) as a function of the top DBR pairs. An initial increase in $\eta$, due to the strengthening Purcell factor $F_P$, is interrupted by a dip caused by strong Bloch mode at $N_\text{Top}=7$, irregardless of $N_\text{Bottom}$ as the mode overlap is determined by the diameter and the path length between the emitter and the topmost interface. Further additions of DBR pairs in the top mirror tune the problematic mode away once more, recovering the high passive efficiency \emph{if} the cavity remains single-sided. If there aren't sufficient pairs in the bottom mirror, the increased reflectivity in the top directs the cavity mode itself into the substrate. A 1-D estimation of the ratio of the Fresnel transmission coefficients $T_\text{Top}/T_\text{Bottom}$, calculated using the Transfer Matrix Method (TMM), is shown in Fig.\ref{BottomDecayFig}(g). The blue region marks upward-facing cavities, pink denotes cavities which point downward, and a white dashed line serves as a visual guide to the “turning” point where further top mirrors start to cause avoidable $\eta$ losses. It can be seen that even for a modest amount of top pairs, $N_\text{Bottom} \geq 30.5$ is needed to optimise the micropillar.

\section{Photon-Phonon Inhibition}

It has long been predicted that the presence of a narrow bandwidth cavity with a significant Purcell enhancement on resonance will reduce coupling to phonon modes by ensuring that the rate of coupling on-resonance to the zero-photon line (ZPL) begins to dominate over emission into the off-resonant phonon sideband (PSB) which extends several meV away from the ZPL even at 0K. However there has been some debate as to whether the Purcell factor required to suppress the PSB to desirable levels means approaching the strong coupling regime~\cite{Iles-Smith2017}. We find here that the situation becomes more promising when moving away from the atomic approximation.  As we have discussed away from resonance from the cavity, a Purcell suppression is observed. As well as reducing decay in undesirable spatial channels, Purcell suppression can be used to reduce the available density of states of the phonon side band (PSB). In contrast to the atomic-like cavity approximation, the micropillar \emph{suppresses} a significant fraction of the PSB due to the significant off-resonant Purcell suppression reducing the off resonant density of states. This is in contrast to the predicted effect of the cavity acting as a passive spectral filter, which simply reduces the overall efficiency.  

    \begin{figure}[h!]
    	\center
    	\includegraphics*[width=0.5\linewidth]{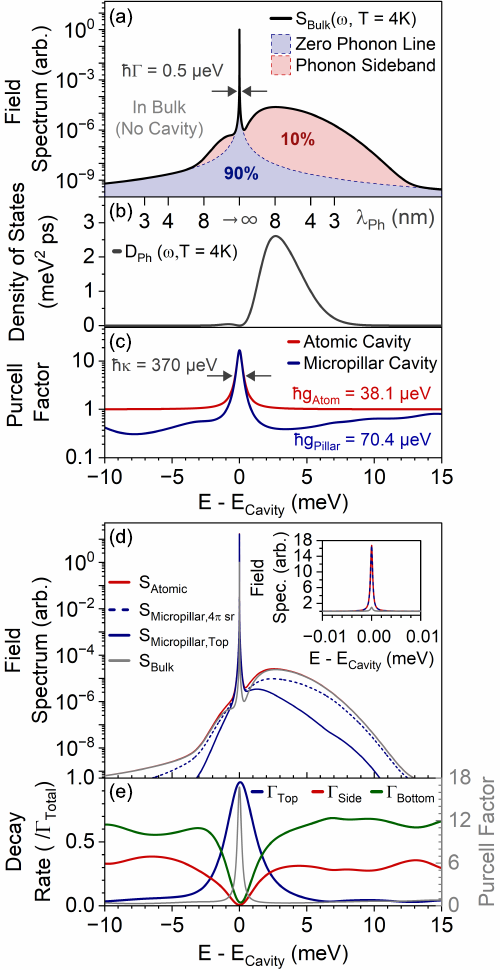}
    	\caption{(a) A simulated field spectrum in bulk GaAs (without any cavity Purcell effects) with a ZPL weight of 90\%. (b) The phonon density of states $D_\text{Ph}$ at 4K (Eq. \eqref{PhononDOSEq}) generates the phonon sideband in the field spectrum. (c) The FDTD-calculated Purcell factor of the 13/35.5 micropillar cavity (blue) shows off-resonant suppression of photon-phonon interactions as $\Gamma_\text{Side} < \Gamma_0$ at large $E-E_\text{Cavity}$, whereas an equivalent atomic emitter would not. (d) The emitted field spectrum in bulk (grey dashed line) is compared to the atomic cavity field (red) where enhancement of the ZPL leads to $W_\text{Atom} = 99.27\%$, and to the micropillar cavity spectrum (blue) where the additional PSB suppression leads to $W_\text{Pillar,$4\pi$ sr} = 99.70\%$, leading to $W_\text{Pillar,Top} = 99.92\%$ in the field which decays through the top and is "filtered" by the cavity. (e) The decay rates through the micropillar faces, normalised to the total power emitted. The suppression of undesired decay channels near the cavity leads to a much wider bandwidth of the efficiency i.e. the width of $\Gamma_\text{Top}/\Gamma_0$ in blue ($=\xi$) is several-fold broader than the width $\kappa$ of the Purcell factor (grey, right axis).}
    	\label{PhononsFig}
	\end{figure}

We quantitatively examine whether the concurrent strengthening of the ZPL and weakening of the PSB field components can be used to near-eliminate any dephasing due to phonons by firstly considering the spectrum of the emitted field in bulk GaAs by a quantum dot on resonance with the cavity and linewidth $\hbar\Gamma_0 = 0.5~\mu$eV as an estimate, which has been shown in Fig\ref{BestPillarsFig}(a). The spectrum $S_\text{Bulk}(\omega)$ consists of a narrow ZPL Lorentzian and a broadband PSB such that the ZPL weight over $4\pi~sr$ is calculated as

\begin{equation}
W_{4\pi} = \frac{\int_{-\infty}^{\infty}S_\text{ZPL}(\omega)d\omega}{\int_{-\infty}^{\infty}S_\text{Bulk}(\omega)d\omega}
\end{equation}

where $S_\text{ZPL}$ is the ZPL portion of the field spectrum. We generate the phonon sideband (PSB) such that $W_{4\pi} = 90\%$ , which errs on the pessimistic side of experimental measurements and theoretical models, where at 4K $W_{4\pi} = 90-93\%$ \cite{Borri2001,Muljarov2004,Mork2019,Iles-Smith2017,Borri2003,Borri2005,Michler2011}. This remaining 10\% of the spectrum comes from the phonon density of states, $D_\text{Ph}(\omega,T)$ (Fig.\ref{BestPillarsFig}(b)), which depends on the vertical and lateral sizes of the ground and excited electron wave functions $L_{i,j}$ where $i \in \{e,g\}$, $j \in \{xy,z\}$ and is given by the analytical and experimental models presented in \cite{Lodahl2013,Lodahl2013Theory},

\begin{equation}
\begin{split}
D_\text{PH}(\omega,T) = \frac{\hbar}{4\pi d c_l^5} &\frac{\omega^3}{1-e^{\frac{-\hbar\omega}{k_\text{B}T}}} \int_0^1 \left( D_e e^{\tilde{\omega}^2_\text{e,xy}\left(u^2-1\right)-\tilde{\omega}^2_\text{e,z}u^2} - D_g e^{\tilde{\omega}^2_\text{g,xy}\left(u^2-1\right)-\tilde{\omega}^2_\text{g,z}u^2}\right) du
\label{PhononDOSEq}
\end{split}
\end{equation}
where $\tilde{\omega}_{i,j} = \omega L_{i,j}/(2 c_L)$, with deformation potentials $D_e = -14.6$ eV and $D_g = -4.8$ eV, and crystal density $d=5370$ kg/m$^3$. The characteristic wavefunction lengths have also been taken from roughly $0.95~\mu m$ dots measured in \cite{Lodahl2013}; as the electron wavefunctions for O-band QDs will be larger, which leads to weaker photon-phonon dephasing \cite{Mork2019}, we are slightly overestimating the phonon contributions to the emitted spectrum and any ZPL weights calculated in this manner represent a lower bound; in reality the phonon sideband could contribute even less. The field spectrum $S_\text{Bulk}$ is then acted upon by the FDTD-calculated cavity. An example of the 13/35.5 $5\lambda/n$ pillar has been plotted in Fig. \ref{BestPillarsFig}(c) (blue line), which is compared to the effects of a similar structure with the atomic cavity approximation, which gives the same peak and FWHM, but a constant $F_P \to 1$ at large detuning. It is worth noting that because the actual micropillar cavity goes to $F_P \to 0.3$, the base-to-peak contrast is much larger than in the atomic case. The suppression of alternative decay leads to stronger coupling as a Lorentzian maximum-to-minimum ratio of $\approx 55$, where the Purcell factor as a function of detuning is given by 

\begin{equation}
    F_P = \frac{\Gamma_\text{Total}}{\Gamma_0} = \Gamma_0 \left(1+2\frac{g^2}{\kappa \Gamma_0} \frac{\frac{\kappa}{2}}{(\frac{\kappa}{2})^2 + \Delta^2}\right)
\end{equation}

where $\Delta$ is the atom-cavity detuning $\omega_\text{Cavity} - \omega_\text{Emitter}$ \cite{Carmichael2009}, we find that $\hbar g_\text{Pillar} = 70.4$ µeV, while $\hbar g_\text{Atom} = 38.1$ µeV; this stronger coupling to the cavity further weakens the dephasing due to phonons \cite{Kaer2013}. The off-resonant Purcell suppression of the field spectrum, as well as the cavity enhancement of decay into the ZPL work simultaneously to bring up the ZPL weight of the detected field (assuming the detector collects only $\Gamma_\text{Top}$) from $W_\text{4$\pi$} > 98\%$ to $W_\text{Top}>99\%$. The spectra shown in \ref{BestPillarsFig}(d), show the enhancement of the ZPL provided by both atomic-like and micropillar cavities in the inset, where the actual micropillar cavity allows for off resonant suppression, resulting in 30x suppression of the PSB. 

Part of the phonon sideband decays preferentially through the side channels, causing some efficiency losses. There have been works in recent years which apply the atomic-cavity estimation to micropillars, concluding that due to overlap between in-cavity photon states and high energy phonons, even a perfect structure ($\xi=1$) incurs unavoidable losses due to the cavity "filtering" the emitted field by redirecting photons  into lossy side channels \cite{Mork2019,Iles-Smith2017}. We examine whether this is the case for our cavities by comparing the emitted spectrum over $4\pi$ steradians with the field decaying through $\Gamma_\text{Top}$. The proportion of the sideband which is not redirected into the side corresponds to $\Gamma_\text{Top}/\Gamma_\text{Total}$ and has been shown in Fig.\ref{BestPillarsFig}(e) (blue, left axis) alongside the cavity Purcell factor (grey, right axis). Whereas it might seem intuitive to treat the cavity linewidth $\kappa$ as the filter FWHM, due to reduction of leaky decay as $E-E_\text{Cavity} \to 0$, the bandwidth of the outcoupling efficiency $\xi_\text{FWHM}$ is in fact several times larger than the cavity linewidth $\kappa$. The results in Fig.\ref{PhononDOSEq} show that the PSB can be reduced by an order of magnitude or more, as we discuss in more detail in the next section.

\section{Bright Pillar Designs and Discussion}

\begin{figure}[hb!]
\centering
\includegraphics[width=0.5\linewidth]{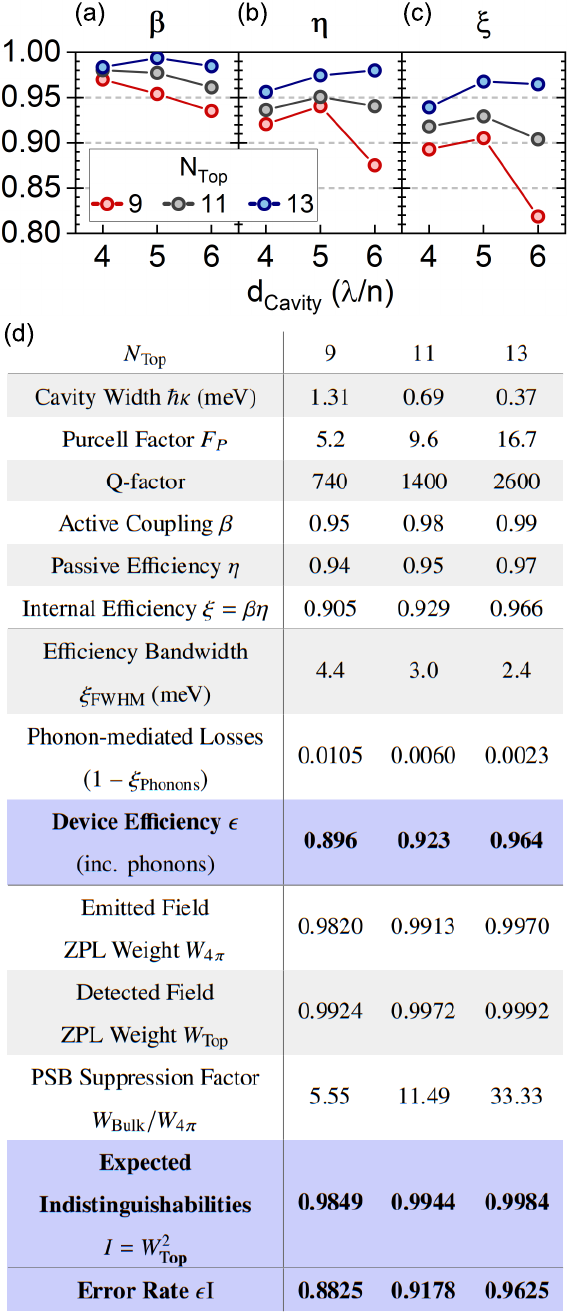}
        \caption{(a-c) The efficiency factors of micropillars at full-wavelength diameters, with $N_\text{Bottom} = 35.5$ and increasing $N_\text{Top}$, identify an optimal $d_\text{Cavity}=5 \lambda/n$. (d) Figures of merit for high-efficiency low-Q micropillars with $N_\text{Bottom}=35.5$ DBR pairs and pillar diameter $d_\text{Cavity} = 5\lambda/n_\text{Cavity}$. Some of the emission into the PSB gets filtered by the collection function $\Gamma_\text{Top}/\Gamma_\text{Total}$, displayed as a blue line in Fig. \ref{PhononsFig}(e). The funnelling parameter $F$ is the fraction of the field not redirected by the cavity. The device efficiency $\epsilon = \xi F$ is the probability of a photon decaying through the top, when considering both cavity and phonon effects on alternate decay channels. As the ZPL weight in the detected field, $W_\text{Top}$, represents the probability of a photon in the ZPL, the indistinguishability can be estimated as $I = W_\text{Top}^2$, or the probability of 2 photons decaying through the ZPL as they would be distinguishable if one or both interact with a phonon.}
        \label{BestPillarsFig}
\end{figure}

After dealing with the major source of losses (side leakage and Bloch modes) and also considering the suppression of the PSB, we are now in a position to propose micropillar designs that simultaneously meet the requirements of (i) very high overall efficiency (ii) very low emission into the phonon sideband and (iii) low enough Q-factor to allow ultrafast pulses of ps to be input. We therefore scan for optimal pillars across several optimal $d_\text{Cavity}$, where $N_\text{Bottom} = 35.5$ ensures a strongly emission out of the top of the pillar and $N_\text{Top} \in \{9,10,11\}$ results in strong yet sufficiently broadband cavities. The on-resonance, peak values of the active coupling, passive efficiency, and the internal efficiency of the cavity are presented in Fig. \ref{BestPillarsFig}(a-c). Following only these two simple steps, we have reduced the dimensionality of the $d_\text{Cavity}$, $N_\text{Top}$, and $N_\text{Bottom}$ optimisation problem from a 3 dimensional space to a few points, and we present designs which all but eliminate the side losses ($\beta = 99\%$ of the 13/35.5 pillar at 5$\lambda/n$ diameter). In examining Bloch mode losses, we note a substantial increase in the $\eta$-factor of the $6\lambda/n$ diameter pillars in Fig.\ref{BestPillarsFig}(b) when $N_\text{Top}$ is incremented. This is the result of a $\Gamma_\text{Bottom}$ propagating Bloch mode being tuned away from the cavity. It is imperative to minimise any spectral overlap, and the $5 \lambda/n$ diameter seems the most suitable choice as large changes in $N_\text{Top}$ result in minimal $\eta$-changes and conversely, $6 \lambda/n$ is unsuitable. We have therefore 3 high performing pillars with $N_\text{Bottom}=35.5$, $N_\text{Top} = \{9,11,13\}$ and efficiency $\xi = \{0.905,0.929,0.966\}$, with respective cavity FWHM of $\approx\{1.31,0.69,0.37\}$ meV.  In Fig.\ref{BestPillarsFig}(a-c) we demonstrate that by including the full 3-D effects of the micropillar structure, the internal efficiency extends well beyond the atomic-like cavity predictions and much closer to near-unity efficiency than previously thought, with each of the 3 designs having $\xi>90\%$ of emitted photons exiting through the top of the micropillar.

Having identified good micropillar designs, we present their figures of merit in Fig. \ref{BestPillarsFig}(d). The on-resonant values include standard cQED parameters, such as the cavity linewidth $\kappa$, the Purcell factor $F_P$, and the Q-factor, as well as the efficiency factors of the cavity. We also report phonon-mediated losses to be at least an order of magnitude lower than in bulk. We estimate a PSB-adjusted device efficiency $\epsilon$ where

\begin{align}
 \xi_\text{Phonon} &= \frac{\int_{-\infty}^{\infty}S_\text{Top}(\omega)d\omega}{\xi\int_{-\infty}^{\infty}S_{4\pi}(\omega)d\omega} \\
\epsilon &= \xi \ \xi_\text{Phonon}
\end{align}

which takes into account losses due to coupling to phonons, that are not included in the FDTD simulations directly. Due to the near-elimination of the phonon sideband, we also report high expected indistinguishabilities $I$, with spectrally stronger cavities giving better values for $\epsilon I$, albeit at the cost of a narrowing resonance. As such, the most suitable design depends on the device, and the pillars presented can fulfil different roles: for instance, a lower overall efficiency but broad micropillar (9/35.5) is more favourable for collection of a biexction-exciton pair, whereas a more narrowband pillar with a higher efficiency (13/35.5) is better suited for efficient generation of high fidelity linear cluster states: $\epsilon I = 0.9625$ leads to a 54\% probability of generating a 16-qubit chain. The 13/35.5 design is also suitable for applications such as boson sampling due to the high indistinguishability, or for use as a light-matter interface in non-linear interactions as a switch or a quantum memory. One should note that for non-linear interactions it is the $W_{4\pi}$ value that is important, unlike for single photon sources where one observes indistinguishabilities of $W_\text{Top}$. On the other hand, coherent control of the quantum dot requires a $\approx 1~ps$ $\pi$-pulse which cannot spectrally fit in the narrow-cavity, making a 9 or 10/35.5 pillar with a $5\lambda/n$ diameter to be more suitable. As shorter pulses reduce the probability of multiphoton absorption events, better purity can be achieved with the low-Q pillar candidates.

While the device efficiency $\epsilon$ represents the brightness at the first lens, the far-field angular distribution of the emitted mode and the corresponding collection efficiency within some NA have been unaddressed in this work. Many of the figures of merit presented here may be improved by various means. Overall extraction efficiencies may be tuned by, for example, introducing adiabatically varying DBR layers and controlling the refractive index contrast of the final top layer of material, and there is similar scope for engineering the Bloch modes to be minimal. Reduction of the coupling to phonon modes may be controlled further by controlling the QD growth or material. We also note that many practical issues still remain to be resolved, some of which are known (spectral wander from charge noise for example, or an examination of fabrication tolerances and deviations from our simple design).  Practical realisation of QD photonic structures with such figures of merit will likely reveal other unanticipated sources of loss and distinguishability, however, we believe that the demonstration of micropillar designs with $\geq 95\%$  overall single photon emission efficiency and suppression of the phonon coupling to give indistinguishabilities of $\geq 99.9\%$  in some cases paves the way for an engineering approach to reaching even higher values.  

\section{Conclusion}

We demonstrate in this work designs for simple micropillar structures using industry standard techniques with record-breaking efficiency in the weak-coupling Purcell regime. We achieve simultaneously in one device design overall single photon extraction efficiency of  $\xi > 97\%$, phonon sideband suppression that should yield photon indistinguishabilities of $I > 99.9\%$ in a simple to fabricate structure with a Q-factor of 2500, low enough to allow optical coherent control pulses of a few ps, and to allow polarization denegerate emission if required. While design of outcoupling strategies goes beyond the scope of this work, we believe our cavity design strategy demonstrates the plausibility of easy-to-fabricate broadband micropillars as near-perfect on-demand sources of indistinguishable photons and deterministic non-linearities for even the most demanding applications, such as fault-tolerant photonic quantum computing.  We emphasise that the figures of merit presented do not represent a theoretical limit, and we anticipate that with careful engineering even higher efficiencies and phonon suppression can be designed. Our designs can also be applied to a wide range of target wavelengths. Moreover, the principles applied to these QD micropillar cavities, i.e. careful consideration of the photonic density of states in all three dimensions, careful understanding of how the cavity can modify the phononic modes, can be applied to other photonic structures. For instance, photonic crystal waveguides could be modified to include low Q-factor adiabatic cavities which would better control the phonon sideband. Likewise, similar Bloch-like leaky modes may appear in bullseye structures, and careful engineering of the phononic density of states with the cavity will be important for other emitters such as colour centres in diamond and SiC. 

\section{Acknowledgements}

The authors would like to acknowledge contributions and useful discussions from Sam Mister and Asciah M.S. Alshahrani. This project has benefited from funding by EPSRC grant EP/N003381/1, and from the Ministry of Education and Science of the Republic of North Macedonia. PA acknowledges financial support provided by EPSRC via Grant No. EP/T001062/1.

\bibliography{Bibliography}

\end{document}